\documentclass{ws-p9-75x6-50}

\def\CMP{\em Commun. Math. Phys.}
\def\JMP{\em J. Math. Phys.}
\def\JSP{\em J. Stat. Phys.}
\def\PRL{\em Phys. Rev. Lett.}

\def\Journal#1#2#3#4{{#1}, #2 (#3), #4}
\def\LL{{\cal L}}
\def\eg{{\sl e.g.\ }}
\def\ie{{\sl i.e.\ }}
\def\bq{{\bf q}}
\def\bp{{\bf p}}
\def\bv{{\bf v}}
\def\br{{\bf r}}
\def\bx{{\bf x}}
\def\bi{{\bf i}}
\def\bj{{\bf j}}
\def\VV{{\cal V}}
\font\msytw=msbm10 scaled\magstep1
\def\RRR{\hbox{\msytw R}}
\def\ZZZ{\hbox{\msytw Z}}

\begin{document}

\title{Fourier Law: A Challenge to Theorists}

\author{F.~Bonetto}

\address{Department of Mathematics, Rutgers University,
110 Frelinghuysen Road,\\
Piscataway NJ 08854.\\
Present address: IHES, 75 route de Chartres, 91440 Bures sur Yvette,
France\\email: bonetto@ihes.fr
}

\author{J.~L.~Lebowitz}

\address{Department of Mathematics and Physics, Rutgers University,
110 Frelinghuysen Road, \\Piscataway NJ 08854.\\
Present address: IHES, 75 route de Chartres, 91440 Bures sur Yvette,
France\\email: lebowitz@ihes.fr
}

\author{L.~Rey-Bellet}
\address{Department of Mathematics, Rutgers University,
110 Frelinghuysen Road, \\Piscataway NJ 08854.\\
Present address: Department of Mathematics, University of Virginia, 
Kerchof Hall, Charlottesville VA 22903\\
email: lr7q@virginia.edu}


\maketitle

\abstracts {We present a selective
overview of the current state of our knowledge (more precisely of our
ignorance) regarding the derivation of Fourier's Law, ${\bf J}(\br) =
-\kappa {\bf \nabla}T(\br)$; ${\bf J}$ the heat flux, $T$ the
temperature and $\kappa$, the heat conductivity.  This law is
empirically well tested for both fluids and crystals, when the
temperature varies slowly on the microscopic scale, with $\kappa$ an
intrinsic property which depends only on the system's equilibrium
parameters, such as the local temperature and density.  There is
however at present no rigorous mathematical derivation of Fourier's
law and ipso facto of Kubo's formula for $\kappa$, involving integrals
over equilibrium time correlations, for any system (or model) with a
deterministic, \eg Hamiltonian, microscopic evolution.}

\section{Introduction}\label{sec:intro}

There are at least two distinct situations in which Fourier's Law is
observed to hold with high precision:
\begin{enumerate}

\item
An isolated macroscopic system which is prepared at some initial time,
say $t=0$, with a nonuniform temperature $T_0({\bf r})$, \eg a fluid
or solid in a domain $\Lambda$ surrounded by effectively adiabatic
walls. 
%This will correspond to a local energy density $e_0(T_0({\bfr})$.  
At $t>0$, the temperature will change, due to the heat, \ie energy,
current, with the energy density satisfying the conservation equation:

\begin{equation}
c_v(T) {\partial \over \partial t} T(\br,t) = 
-\nabla \cdot {\bf J} = \nabla \cdot \left[\kappa \nabla T\right], 
\label{1.1}
\end{equation}
where $c_v(T)$ is the specific heat per unit volume and we have
assumed that there is no mass flow or other mode of energy transport
beside heat conduction (we also ignore for simplicity any variations
in density or pressure).  Eq.(\ref{1.1}) is to be solved subject to the
initial condition $T(\br,0) = T_0(\br)$ and no heat flux
across the boundary of $\Lambda$.  The stationary state, achieved as
$t \to \infty$, is then one of uniform temperature $\bar T$ determined
by the constancy of the total energy. For our purposes we can also
think of $\Lambda$ as a torus, \ie having periodic boundary conditions.

\item
We consider the system in contact with heat reservoirs which specify a
time invariant temperature $T_\alpha$ at points of the boundary
$\br\in (\partial \Lambda)_\alpha$ in contact with the $\alpha$-th heat
reservoir, $\alpha\geq 1$.  When the system has come to a stationary
state (again assuming no matter flow) its temperature will be given by
the solution of Eq.(\ref{1.1}) with the left side set equal to zero,
\begin{equation}
\nabla \cdot \tilde J(\br)=
\nabla \cdot (\kappa \nabla \tilde T(\br)) = 0 \label{1.2},
\end{equation}
subject to the boundary condition $\tilde T(\br) = T_\alpha$ for $\br
\in (\partial\Lambda)_\alpha$ and no flux across the rest of the
boundary which is insulating or periodic in the direction
perpendicular to the heat flow. A simple example of this situation is
the usual set up for a Benard experiment in which the top and bottom
of a fluid in a cylindrical slab of height $h$ and cross sectional
area $A$ are kept at different temperatures $T_h$ and $T_b$
respectively. (To avoid convection one has to make $T_h>T_b$ or keep
$|T_h-T_b|$ small). Assuming uniformity in the direction parallel to
the vertical $x$-axis one has in the stationary state a temperature
profile $\tilde T(x)$ with $\tilde T(0) = T_b$, $\tilde T(h) = T_h$
and $\kappa(\tilde T) {d \tilde T \over dx} =$Const. for $x \in
(0,h)$.

\end{enumerate}

From a physical point of view, which is how we presented them, the two
cases are conceptually very similar (some physicists would even say
identical).  We have implicitly assumed that the system is described
fully by specifying its temperature $T(\br,t)$ everywhere in
$\Lambda$.  What this means on the microscopic level is that we
imagine the system to be in local thermal equilibrium (LTE).  To make
this a bit more precise we might think of the system as being divided
up (mentally) into many little cubes, each big enough to contain very
many atoms yet small enough on the macroscopic scale to be accurately
described, at a specified time $t$, as a system in equilibrium at
temperature $T(\br_i,t)$, where $\br_i$ is the center of the $i$-th
cube. For slow variation in space and time we can then use a
continuous description $T(\br,t)$.

This notion is made precise in the so called hydrodynamic scaling
limit (HSL) where the ratio of micro to macro scale goes to
zero\cite{KiLa,Sp,LePrSp}. The macroscopic coordinates ${\bf r}$ and
$t$ are related to the microscopic ones $\bq$ and $\tau$, by
$\br=\epsilon\bq$ and $t =\epsilon^\alpha\tau$, \ie if $\Lambda$ is a
cube of macroscopic sides $l$, then its sides, now measured in
microscopic length units, are of length $L = \epsilon^{-1} l$. We then
suppose that at $t=0$ our system of $N=\rho L^d$ particles with
Hamiltonian
\begin{equation}
H(P,Q)=\sum_{i=1}^{N}\left[\frac{\bp_i^2}{2m}+\sum_{j\not=i}\phi(\bq_j-\bq_i)
+ u(\bq_i)\right]=\sum_{i=1}^{N}\frac{\bp_i^2}{2m}+{\cal V}(Q)
\label{ham_f}
\end{equation}
is described by an equilibrium Gibbs measure with a temperature
$T(\br) = T(\epsilon\bq)$: roughly speaking the phase space ensemble
density has the form,
\begin{equation}
\mu_0(P,Q) \sim \exp\left\{-\sum_{i=1}^N \beta_0(\epsilon \bq_i)
\left[{{\bp_i}^2 \over 2m} + \sum_{j\ne i} \phi(\bq_j-\bq_i) +
u(\bq_i)\right]\right\}\,,
\label{loceq} 
\end{equation}
where $Q=(\bq_1,\ldots,\bq_N)\in\Lambda^{dN}$,
$P=(\bp_1,\ldots,\bp_N)\in\RRR^{dN}$, $\phi(\bq)$ is some short range
inter particle potential, $u(\bq_i)$ an external potential and
$\beta_0^{-1}(\br)= T_0(\br)$\cite{Sp}.

In the limit $\epsilon \to 0$, $\rho$ fixed, the system at $t=0$ will
be macroscopically in LTE with a local temperature $T_0(\br)$ (as
already noted we suppress here the variation in the particle density
$n({\bf r})$).  We are interested in the behavior of a macroscopic
system, for which $\epsilon<<1$, at macroscopic times $t\geq 0$,
corresponding to microscopic times $\tau=\epsilon^{-\alpha} t$,
$\alpha = 2$ for heat conduction or other diffusive behavior.  The
implicit assumption then made in the macroscopic description given
earlier is that since the variations in $T_0({\bf r})$ are of order
$\epsilon$ on a microscopic scale, then for $\epsilon << 1$, the
system will, also at time $t$, be in a state very close to LTE with a
temperature $T({\bf r},t)$ that evolves in time according to Fourier's
law, Eq.(\ref{1.1}).

>From a mathematical point of view the difficult problem is proving
that the system stays in LTE for $t>0$ when the dynamics are given by
a Hamiltonian time evolution. This requires proving that the
macroscopic system has some very strong ergodic properties, \eg that
the only time invariant measures locally absolutely continuous w.r.t
Lebesgue measure are, for infinitely extended spatially uniform
systems, of the Gibbs type \cite{FFL,OVY,LO}.  This has only been
proven so far for systems evolving via stochastic dynamics, \eg
interacting Brownian particles or lattice gases. In these systems the
relevant conserved quantity is usually the particle density rather
than the energy density.  We shall not discuss such stochastic
evolutions here but refer the reader to \citelow{Sp,KiLa} for a
mathematical exposition.

The only Hamiltonian system for which a macroscopic transport law has
been derived is a gas of noninteracting particles moving among a fixed
array of periodic convex scatterers (periodic Lorentz gas or Sinai
billiard). For this system one can prove a diffusion equation like
Eq.(\ref{1.1}) for the density of the particles, both for the initial
value and the suitably defined stationary state problem, with the
(self)diffusion constant given by the Einstein-Green-Kubo formula
\cite{BuSi,LeSp78}. Unfortunately, the absence of interactions between
particles makes this system a poor model for heat conduction in
realistic systems. In particular there is no mechanism for achieving
LTE. The speed of each particle $|\bv|$ does not change in the course
of time and the diffusion constant for each particle is proportional
to its speed. The diffusion equation for the density mentioned above
are therefore in fact separate uncoupled equations for particles with
specified speeds. It corresponds to the usual diffusion equation only
when all the particles have the same speed.

To remedy this problem it would be necessary to add interactions
between the moving particles, \eg instead of points make them little
balls, and then derive coupled equations for the diffusion of both
particle and energy densities.  This is what we would consider a
satisfactory answer to the challenge in the title of this article and
we offer a bottle of very good wine to anyone who provides it. We
believe that this system, with only two conservation laws and an
external source (the fixed convex scatterers) for chaotic dynamics may
be the simplest Hamiltonian system for which such results could be
proven rigorously. 

Just how far we are from such results will become clear as we describe
our current mathematical understanding of the stationary
nonequilibrium state (SNS) of macroscopic systems whose ends are, as
in the example of the Benard problem, kept at fixed temperatures
$T_1$ and $T_2$.  The heat conductivity in this situation can be
defined precisely without invoking LTE.  To do this we let $\tilde J$
be the expectation value in the SNS, \ie we assume that the SNS is
described by a phase-space measure (whose existence we discuss later),
of the energy or heat current flowing from reservoir 1 to reservoir 2.
We then define the conductivity $\kappa_L$ as $\tilde J/(A\delta T/L)$
where $\delta T/L= (T_1-T_2)/L$ is the effective temperature gradient
for a cylinder of microscopic length $L$ and uniform crossection $A$
and $\kappa(T)$ as the limit of $\kappa_L$ when $\delta T\to 0$
($T_1=T_2=T$) and $L\to \infty$\cite{Le2}.  The existence of such a
limit with $\kappa$ positive and finite is what one would like to prove.

\section{Heat Conduction in Gases}\label{sec:Kinth}

Before going on to a mathematical discussion of heat conducting SNS,
we turn briefly to the ``kinetic theory'' analysis of heat conduction
in gases.  This is historically the first example of a microscopic
description of this macroscopic phenomenon.  It goes back to the works
of Clausius, Maxwell and Boltzmann \cite{Brush} who obtained a
theoretical expression for the heat conductivity of gases, $\kappa
\sim\sqrt{T}$, independent of the gas density. This agrees with
experiment (when the density is not too high) and was a major early
achievement of the atomic theory of matter\cite{Brush}.

Clausius and Maxwell used the concept of a ``mean free path''
$\lambda$: the average distance a particle (atom or molecule) travels
between collisions in a gas with particle density $\rho$.
Straightforward analysis gives $\lambda \sim 1/\rho \pi \sigma^2$,
$\sigma \sim$ an ``effective'' hard core diameter of a particle. They
considered a gas with temperature gradient in the $x$-direction and
assumed that the gas is (approximately) in local equilibrium with
density $\rho$ and temperature $T(x)$. Between collisions a particle
moves a distance $\lambda$ carrying a kinetic energy proportional to
$T(x)$ from $x$ to $x+\lambda/\sqrt{3}$, while in the opposite
direction the amount carried is proportional to $T(x + \lambda\sqrt
3)$.  Taking into account the fact that the speed is proportional to
$\sqrt{T}$ the amount of energy transported per unit area and time
across a plane perpendicular to the $x$-axis $J$ is approximately,
\begin{equation}\label{cond}
J \sim \rho\sqrt{T}\left[T(x)-T(x+\lambda\sqrt 3)\right]\simeq
-\sigma^{-2}\sqrt{T}\frac{dT}{dx} \,,
\end{equation}
and so $\kappa \sim \sqrt {T}$ independent of $\rho$, in agreement
with experiment.  

It was clear to the founding fathers that starting with a local
equilibrium situation (corresponding to a Maxwellian distribution of
velocities) there will develop, as time goes on, a deviation from
LTE. They reasoned however that this deviation from local equilibrium
will be small when $(\lambda/T)dT/dx<< 1$, the regime in which
Fourier's law is expected to hold, and the above calculation should
yield, up to some factor of order unity, the right heat
conductivity. In fact if one computes the heat flux at a point $x$ by
averaging the microscopic energy current at $x$ $j=\rho\bv
(\frac{1}{2}mv^2)$ over the one particle distribution function
$f(\br,\bv,t)$ then it is only the deviation from local equilibrium
which makes a contribution. The result however is essentially the same
as Eq.(\ref{cond}). This was shown by Boltzmann who derived an
accurate formula for $\kappa$ in gases by using the Boltzmann equation
to compute $\kappa$. If one takes $\kappa$ from experiment the above
analysis yields a value for $\sigma$, the effective size of an atom or
molecule, which turns out to be close to other determinations of the
characteristic size of an atom\cite{Brush}. This gave evidence for the
reality of atoms and the molecular theory of heat.

Using ideas of hydrodynamical space and time scaling described earlier
it is possible to derive a controlled expansion for the solution of
the stationary Boltzmann equation describing the steady state of a gas
coupled to temperature reservoirs at the top and
bottom\cite{ELM1,ELM2,ELM3}. The coupling is implemented by the
imposition of ``Maxwell boundary conditions'': when a particle hits
the left (right) wall it get reflected with a distribution of
velocities
\begin{equation}
f_{\alpha}(d\bv)=\frac{m^2}{2\pi(kT_\alpha)^2}
|v_x|\exp \left[ -\frac{m \bv^2}{2kT_\alpha}\right]d\bv\qquad\alpha=1,2
\end{equation}
corresponding to a temperature $T_1$ ($T_2$) at the left (right) wall.
One then shows\cite{ELM1,ELM2,ELM3} that for $\epsilon<<1$, $\epsilon$
being now the ratio $\lambda/L$, the Boltzmann equation for $f$ in the
slab has a time independent solution which is close to a local
Maxwellian, corresponding to LTE, (apart from boundary layer terms)
with a local temperature and density given by the solution of the
Navier-Stokes equations which incorporates Fourier's law as expressed
in Eq.(\ref{1.2}). The main mathematical problem is in controlling the
remainder in an asymptotic expansion of $f$ in power of
$\epsilon$. This requires that the macroscopic temperature gradient,
\ie $|T_1-T_2|/h$, where $h=\epsilon L$ is the thickness of the slab
on the macroscopic scale, be small.

Even if this apparently technical problem could be overcome we would
still be left with the question of justifying the Boltzmann equation
for such steady states and of course it would not tell us anything
about dense fluids or crystals. In fact the Boltzmann equation itself
is really closer to a macroscopic then to a microscopic description.
It is obtained in a well defined kinetic scaling limit in which in
addition to rescaling space and time the particle density goes to
zero\cite{Sp}, \ie $\lambda>>\sigma$.

\section{Heat conduction in insulating crystals}\label{sec:heat}

Excellent accounts of the historical development of the theory of heat
conduction in solids exist~\cite{Pe2,Ku} so we will content ourselves
here with some brief remarks. In (electrically) insulating solids,
heat is transmitted through the vibrations of the lattice (in
conductors the electronic contribution is in general much larger then
the contribution due to the lattice vibrations).  In order to use
concepts of kinetic theory, it is useful to picture a solid as a gas
of phonons which can store and transmit heat. In a perfectly harmonic
crystal, the phonons behave like a gas of noninteracting particles and
therefore the thermal current will not decrease with the length of the
crystal placed between two thermal reservoirs. Thus a perfectly
harmonic crystal has an infinite thermal conductivity: in the language
of kinetic theory $\sigma=0$ and the mean free path $\lambda$ is
infinite. A real crystal is not harmonic and, in the phonon picture,
any thermal current will be degraded by the anharmonic forces in the
lattice. Another source of finite thermal conductivity may be the
lattice imperfections and impurities which will scatter the phonons
and degrade the thermal current too.

Debye~\cite{De} devised a kind of kinetic theory for phonons in order
to describe thermal conductivity. One assumes that a small gradient of
temperature is imposed and that the collisions between phonons
maintain local equilibrium. An elementary argument~\cite{AsMe} gives a
thermal conductivity analogous to Eq.(\ref{cond}) obtained in Section
\ref{sec:Kinth} for gases, (remembering however that the density of
phonon is itself a function of $T$)
\begin{equation}\label{cond1}
\kappa \sim  c_v c^2 \tau\,.
\end{equation}
Comparing Eq.(\ref{cond1}) and Eq.(\ref{cond}) we see that $\rho$ has
been replaced by $c_v$ the specific heat of the phonons, $\sqrt{T}$ by
$c$ the (mean) velocity of the phonons, and $\lambda$ by $c\tau$,
where $\tau$ is the effective mean free time between phonon
collisions. The thermal conductivity depends on the temperature via
$\tau$ and a more refined theory is needed to account for this
dependence. Peierls~\cite{Pe1} used a Boltzmann type equation for
phonons to investigate this problem. The Peierls theory singles out
one phenomenon which gives rise to a finite thermal
conductivity\cite{Pe2}. The momentum of phonons in collisions is
conserved only modulo a vector of the reciprocal lattice. One can
therefore classify the collisions of phonons into two classes: the
ones where phonon momentum is conserved (the normal processes) and the
ones where the initial and final momenta differ by a non-zero
reciprocal lattice vector (the umklap processes). Peierls theory may
be summarized (very roughly) as follows: in the absence of umklap
processes the mean free path and thus the thermal conductivity of an
insulating solid is infinite.

A success of Peierls theory is to describe correctly the temperature
dependence of the thermal conductivity~\cite{AsMe}.  Furthermore, on
the basis of this theory, one does not expect a finite thermal
conductivity in $1$-dimensional mono-atomic lattices with pair
interactions: this seems so far a correct prediction, see Section
\ref{num_res}.

The justification of the Boltzmann equation for phonons has been
questioned~\cite{Ja}. Various alternative mechanisms have been
proposed which would give rise to a finite thermal conductivity, but
it seems fair to say that, so far, no better theory of heat conduction
in insulators has been proposed. As Peierls himself puts
it~\cite{Pe2}: ``It seems there is no problem in modern physics for
which there are on record as many false starts, and as many theories
which overlook some essential feature, as in the problem of the
thermal conductivity of [electrically] non-conducting crystals''.

To find a mathematical description of thermal conduction in crystals
we need to specify the Hamiltonian of the system or at least some
appropriately idealized version of it.  A model crystal is
characterized by the fact that all atoms oscillate around given
equilibrium positions. The equilibrium positions can be thought of as
the points of a regular lattice in $\RRR^d$. For simplicity we will
assume that the lattice is simply $\ZZZ^d$. Although $d=3$ is the
physical situation one can be interested also in the case $d=1,2$.
(The $d=1$ system may show finite thermal conductivity without violating
the Peierls criteria if we admit one particle, non momentum
conserving interactions.)

Let $\Lambda \subset {\ZZZ}^d$ be a finite set and denote by $N$ its
cardinality. Each atom is identified by its position
$\bx_\bi=\bi+\bq_\bi$ where $\bi\in\Lambda$ is the equilibrium
position and $\bq_\bi \in {\bf R}^d$ is the displacement of the
particle at lattice site $\bi$ from this equilibrium position, and we
denote by $\bp_\bi$ its momentum and $m$ its mass.  Since inter
atomic forces in real solids have short range, it is reasonable to
assume that the atoms interact only with their nearest neighbors via a
potential that depends only on the relative distance with respect to
the equilibrium distance.

As already noted it is useful to allow an external confining $1$-body
potential which breaks the translation invariance.  Accordingly the
Hamiltonians that we consider have the general form
\begin{equation}\label{ham} 
H(P,Q)\,=\,\sum_{\bi\in\Lambda} \frac{\bp_\bi^2}{2m}
+ \sum_{|\bi-\bj|=1} V(\bq_\bi-\bq_\bj)+\sum_\bi U_\bi(\bq_\bi)=
\,\sum_{\bi\in\Lambda} \frac{\bp_\bi^2}{2m}
+ {\cal V}(Q)\,,
\end{equation}
where $P=(\bp_\bi)_{\bi\in\Lambda}$ and analogously for $Q$. We shall
further assume that as $|\bq|\to\infty$ so do $U_\bi(\bq)$ and
$V(\bq)$.  The addition of $U_\bi(\bq)$ pins down the crystal and ensures
that $\exp{\left[-\beta H(P,Q)\right]}$ is integrable with respect to
$dPdQ$ and thus the corresponding Gibbs measure is well defined.
Observe that for many purpose it is enough to put the potential
$U_\bi$ on only some of the atoms, \eg the ones on the boundary of
$\Lambda$\cite{RiLeLi,Na}. We note finally that when
$\Lambda\subset\ZZZ^d$ one can still consider that $\bp_\bi$ and
$\bq_\bi\in\RRR^\nu$, $\nu\not=d$, but we will generally assume that
$\bq_\bi\in\RRR^d$.

\noindent{\bf Remark}: While the Hamiltonian in Eq.(\ref{ham}) looks
similar to that in Eq.(\ref{ham_f}) the meaning and domain of the $Q$
variables is entirely different. In a fluid all the particle are
identical and the particle with label $i$ interacts with any
other particle whose position $\bq_j$ is close to $\bq_i$,
$\bq_l\in\Lambda\subset\RRR^d$, $l=1,\ldots,N$. The pair interaction
potential $\phi(\bq)$ is of finite range, \eg hard balls, or decays
rapidly with distance. For the crystal in Eq.(\ref{ham}) $\bq_i$ is
the {\it deviation} from an equilibrium position
$\bi\in\Lambda\subset\ZZZ^d$, etc.

\section{Microscopic models of heat reservoirs}\label{micro}

To produce a stationary heat flow in a system, be it a gas or a
crystal, the system must be coupled to at least two heat reservoirs at
different temperatures.  A physical coupling is one which acts only
at the boundary of the system leaving the dynamics in the bulk purely
Hamiltonian.

Since a realistic description of heat reservoirs and coupling is out
of the question various model reservoirs have been used in analytical
and numerical studies. We give here some examples which will be used
later (other choices are of course possible). The expectation is of
course that the different models will give the same behavior away from
the boundary when the system is macroscopic. This has not been proven
in any example, see \citelow{G1997,G1998}

\subsection{Stochastic reservoirs.}  

We have already discussed one such model of reservoirs commonly used
for fluids in Section \ref{sec:Kinth}. This corresponds to Maxwell
boundary condition discussed in Eq.(\ref{loceq}) for a gas in a
rectangular slab. More generally given a fluid in a domain
$\Lambda\subset\RRR^d$ a particle hitting the wall of the container
confining the system at a point $\br\in\partial\Lambda$ will bounce
back into $\Lambda$ with a Maxwellian distribution of momenta
\begin{equation}\label{max}
f_\br(d\bp)=\frac{\beta(\br)^2}{2\pi m^2}
\bp\cdot\hat n(\br)e^{-\beta(\br)\frac{\bp^2}{2m}}d\bp \,,
\end{equation}
where $\hat n(\br)$ is the inward directed unit vector normal to
$\partial\Lambda$ at $\br$ and $\beta^{-1}(\br)$ is the preassigned
temperature at $\br$.

For solids, which are usually not confined to any fixed spatial region
by external walls, it is sometimes mathematically convenient to use
Langevin type reservoirs which act on the atoms at the ``edge'' of the
crystal.  For definiteness we will choose $\Lambda$ to be a chain of
particles or a parallelepiped in higher dimension (with suitable
boundary conditions), $\Lambda = \{ \bi\in {\ZZZ^d}\,;\, 1\le i_k \le
N_k\,,\, 1\le k \le d \}$.  We assume that the particles at the
``left'' boundary $\{\bi \in \Lambda\,;\, i_1=1\}$ are coupled to a
heat reservoir at temperature $T_L$ and that the particles at the
``right '' boundary $\{\bi \in \Lambda\,;\, i_1=N_1\}$ are coupled to
a heat reservoir at temperature $T_R$. We set $\LL\equiv N_1$
the length of the crystal and $A \equiv N_2 \cdots N_\nu$ its cross
section.

For the particles at the boundary of the crystal in contact with a
heat reservoir, the Hamiltonian equations of motion are modified by
the addition of an Ornstein-Uehlenbeck process:
\begin{eqnarray}
m_\bi{\dot \bp}_{\bi}\,&=&\,-\nabla_{\bq_\bi}\VV(\bq) - 
\lambda_{\alpha} \bp_\bi/m_\bi +
(2\lambda_{\alpha}T_{\alpha})^{1/2} \xi_{\alpha}(t)\,. \label{oe} 
\end{eqnarray}  
In Eq.(\ref{oe}), $\alpha \in \{L,R\}$ are indices of the reservoirs,
$\lambda_\alpha$ describes the strength of the coupling to the
reservoir with temperature $T_\alpha$ of the reservoir, and
$\xi_{\alpha}(t)$ is a white noise, \ie, a Gaussian random processes
with covariance $\left<\xi_{\alpha}(t)\xi_{\beta}(s)\right> =
\delta_{\alpha\beta}\delta(t-s)$.  The form of the coefficients is
chosen so that the dynamics satisfy detailed balance. This implies
in particular that if the system is coupled to a single reservoir at
temperature $T$, then the Gibbs measure with density $Z^{-1}
\exp\left(-T^{-1} H(P,Q)\right)$ is a stationary state of the
system. 

With any such a choice of stochastic process to model the reservoirs
the dynamics is described by a stationary Markov process in the phase
space of the system $\Omega$.

\subsection{Hamiltonian reservoirs.} 

In this case the reservoirs themselves are modeled by infinite
Hamiltonian systems and the full system consisting of
reservoirs+system is Hamiltonian\footnote{ A series of results has
been obtained for systems coupled to a single reservoir, such as
non-relativistic atoms coupled to the quantized electromagnetic field
at temperature zero~\cite{BaFrSi1,BaFrSi2,BaFrSi3}, finite
level atoms coupled to a boson field at positive
temperature~\cite{JaPi1,JaPi2,DeJa}, classical particles coupled to a
scalar field at temperature zero~\cite{KoSp,KoSpKu} or at positive
temperature~\cite{JaPi3}.}. An alternative but equivalent point of
view is to start with an infinite system and to consider a finite
subsystem of it as the system and the remaining part as the
reservoirs. A non-equilibrium situation is obtained by choosing
suitable initial conditions for the part of the total system which
describes the reservoirs, \eg the initial conditions of the reservoirs
are assumed to be distributed according to a Gibbs measure with
corresponding temperatures\footnote{For quantum spin systems axioms are 
formulated in \citelow{Ru99} which establish the existence of a stationary 
state and its mixing property for finite systems coupled to several 
reservoirs at different temperatures}.

Studied in \citelow{RuGr,SpLe,OcLe} the simplest version of such a
total system consists of an infinite chain. The ``left'' reservoir
consists of the particles with labels in $(-\infty, -N]$ and the right
reservoir consists of the particles with labels in $[N, +\infty)$. The
system consists of the particles in the middle. At time $t=0$, the
reservoirs are assumed to be in thermal equilibrium at temperatures
$T_L$ and $T_R$. This is clearly readily generalized to higher
dimensions.

\subsection{Hamiltonian/Stochastic reservoir.}\label{HamSto}

While Hamiltonian reservoirs are in principle the right ones to use
they are totally intractable without further simplifications. When
this is done it is actually possible to find examples in which one
starts with a Hamiltonian reservoir and by integrating out over the
degrees of freedom of the reservoirs, ends up with a stochastic
evolution. Many models of this type have been
constructed\cite{GoLeRa1,GoLeRa2,FaGoSp}. We describe here a model
considered in \citelow{JaPi3,EPR1,EPR2,EH}. The system is a finite
chain of anharmonic oscillators coupled at each end to a reservoir
modeled by a linear $d$-dimensional wave equation, which is the
continuum limit of a $d$-dimensional lattice of harmonic
oscillators. The dynamics of the infinite system, crystal+reservoirs,
is Hamiltonian. One makes the statistical assumption that, at time
$t=0$, the reservoirs are in thermal equilibrium at temperatures $T_L$
and $T_R$. Since the reservoirs are linear, one may integrate them
out, and, by our assumptions on the initial conditions of the
reservoirs, the resulting dynamics for the crystal is stochastic,
though in general not Markovian.  Nevertheless, the fact that the
reservoirs are described by a wave equation, together with special
choices of the coupling between the reservoirs and the chain,
permits~\cite{EPR1} enlarging the phase space of the crystal with a
finite number of auxiliary variables so that the dynamics is Markovian
on the enlarged phase space.

In the simplest case of coupling one variable per reservoir is enough
and the resulting equations for the $N$ oscillators are
\begin{eqnarray}
{\ddot \bq_1} \,&=&\, -\nabla_{\bq_1}\VV(Q) + {\bf r}_L\,, \nonumber\\
{\ddot \bq_j} \,&=&\, -\nabla_{\bq_j}\VV(Q)\,, \quad j=2,\dots,N-1\,,
\nonumber\\ {\ddot \bq_n} \,&=&\, -\nabla_{\bq_n}\VV(Q) + {\bf r}_R\,,
\nonumber\\ {\dot {\bf r}}_L \,&=&\, -\gamma_L ({\bf r}_L -
\lambda_L^2 \bq_1) + (2\gamma_L\lambda_L^2 T_L)^{1/2} {\dot w}_L(t)\,,
\nonumber\\ {\dot {\bf \dot r}}_R \,&=&\, -\gamma_R ({\bf r}_R -
\lambda_R^2 \bq_1) + (2\gamma_R\lambda_R^2 T_R)^{1/2} {\dot w}_R(t)
\,, \label{epreq}
\end{eqnarray}
In Eqs.(\ref{epreq}) $\lambda_L$ and $\lambda_R$ describe the coupling
strength to the reservoirs, $\gamma_L$, $\gamma_R$ are parameters
describing the coupling and ${\dot w}_L$, ${\dot w}_R$ are white
noises. 

If the temperatures of both reservoirs are the same, $T_L=T_R=T$, then
the stationary state is given by the generalized Gibbs measure with
density
\begin{equation}\label{gg}
Z^{-1} \exp{\left(-\frac{1}{T} G(P,Q,R)\right)} 
\end{equation}
where $Z$ is a normalization constant and the generalized
``Hamiltonian'' $G$ is given by
\begin{equation}
G(P,Q,R) \,=\, \left( \frac{\br_L^2}{2\lambda^2_L} - \bq_1 \br_L
\right) + \left( \frac{\br_R^2}{2\lambda^2_R} - \bq_N r_R \right)+
H(P,Q)\,.
\end{equation} 
If one integrates the generalized Gibbs state, Eq.(\ref{gg}), over the
auxiliary variables $\br_L$ and $\br_R$ one finds
\begin{equation}
\int d\br_L d\br_R \, Z^{-1} \exp{\left(-\frac{1}{T} G(P,Q,R)\right)} \,=\,
\tilde Z^{-1} \exp{\left(-\frac{1}{T} H_{\rm eff}(P,Q)\right)}\,,
\end{equation}
where $H_{\rm eff}(P,Q)= H(P,Q)+ \lambda_L^2\bq_1/2+
\lambda_R^2\bq_n/2 $ and $\tilde Z$ a normalization constant.  In view of
this it is natural to consider $H_{\rm eff}(P,Q)$ as the energy of the
chain.

\subsection{Thermostats.}\label{Ther}

A fourth way of modeling the reservoirs is by deterministic
(non-Hamiltonian) forces\cite{EvMo,GaCo}. Such models of reservoirs
are usually called thermostats.  An example of such reservoirs which
are widely used in numerical work, are the so called
Nos\'e-Hoover\cite{No,Ho} thermostats. Imposing these thermostats on
small parts of the system\footnote{It is also possible to give
``mechanized'' models of heat transport in which one imposes the
presence of a heat flow through the application of a particular force
on the bulk of the system. This strong modification of the Hamiltonian
character of the dynamics seems unnatural to us, although it can be
useful for numerical simulations, and we will not discuss these models
here \cite{EvMo,EZ}.}  (on the left and on the right) $\Lambda_L$ and
$\Lambda_R$, the equations of motion of particles in those region of
the box are respectively
\begin{eqnarray}
m{\ddot \bq}_{\bi}\,=& \,-\nabla_{\bq_\bi}\VV(Q) - \zeta_L {\dot
\bq}_\bi \cr
m{\ddot \bq}_{\bi}\,=& \,-\nabla_{\bq_\bi}\VV(Q) - \zeta_R {\dot 
\bq}_\bi \label{nhgr}
\end{eqnarray}
where $\bq_i\in\Lambda_\alpha$ for a fluid and $\bi\in\Lambda_\alpha$
for a crystal, $\alpha\in \{L,R\}$.  The variable $\zeta_\alpha$ model
the action of the thermostat and satisfy the equations
\begin{equation}
{\dot \zeta}_\alpha\,=\, 
\frac{1}{\Theta^2} \left( \frac{1}{T_\alpha}
\sum_{\bi\in\Lambda_\alpha}\frac{\bp_\bi^2}{2 m}-1\right)\,.
\label{nh2}
\end{equation} 
In Eq.(\ref{nh2}), $\Theta$ is interpreted as the response time of the
reservoir and $T_\alpha$ is the temperature of the $\alpha$-th reservoir.

A limiting case of Eqs.(\ref{nhgr})(\ref{nh2}) is when we let
$\Theta\to0$. This limit can be formally taken and the model becomes
equivalent to the so called Gaussian thermostat. This means that one
computes $\zeta_\alpha$ as a function of $P$ and $Q$ in such a way
that the kinetic energy of the particles in $\Lambda_L$ or $\Lambda_R$
is a constant of the motion. After a simple calculation one gets for
the chain:
\begin{equation}
\zeta_L(Q,P)=\frac{\sum_{i<I_L}\bp_\bi
({\bf f}(\bq_\bi-\bq_{\bi+1})-{\bf f}(\bq_{\bi-1}-\bq_\bi))+
{\bf f}_l(\bq_\bi))}{\sum_{i<I_L}\bp_\bi^2}\,,
\end{equation}
and similarly for $\zeta_R(Q,P)$. Here ${\bf f}_l(\bq)=- \nabla U_\bi(\bq)$.
One may also prescribe Gaussian thermostat in which the total
energies, instead of just the kinetic energies, in $\Lambda_L$ and
$\Lambda_R$ are kept fixed.

\section{Existence and Nature of Heat Conducting SNS}\label{rigstate}

Suppose we are given a system described by a Hamiltonian of the form
Eq.(\ref{ham_f}) or Eq.(\ref{ham}) and that we have chosen a given
model of heat reservoirs.  We shall now formulate a sequence of
statements (of increasing mathematical difficulty) on the properties
of the resulting dynamical system.

\subsection{Existence, Uniqueness and approach to the Stationary State.}

The first property that we want to prove is existence and if possible
also uniqueness of a stationary state. For the case when all reservoirs
are at the same temperature $T$ existence is generally obvious -
after all the reservoirs are chosen so that they leave the canonical
Gibbs distribution or some variation of it invariant under the time
evolution. Uniqueness and approach to this equilibrium state presents
more of a problem and may not even be true for certain type of
reservoirs and initial states. 

The real problem of interest for us is when the reservoirs are at
different temperatures. We expect that if the dynamics is stochastic
(\eg for models 4.1 and 4.3 of reservoirs) then ``almost any'' initial
distribution of the state of the system converges to a unique
stationary state which is mixing. This is however, in general, a
mathematically non-trivial problem. The isolated system has a
non-compact phase space and has many invariant states. The coupling to
the reservoirs induces a drift towards a state determined by the
reservoirs.  Since however the coupling to the reservoirs occurs only
at the boundary, the proof of the existence of an invariant measure
requires a good understanding of how energy is transmitted through the
system. There are in fact only few examples (to be discussed in
Section \ref{exa}) where this behavior has been proven.

For general Hamiltonian or thermostated reservoirs the problem seems
to be mathematically out of reach at the present time. Starting in a
state that corresponds to the product of two equilibrium states for
the reservoirs times a generic initial distribution for the crystal we
then expect, that in the long time limit the marginal distribution for
the system will approach some limit. When the two infinite reservoirs
are initially at different temperatures, the limiting state should
describe a system having a temperature gradient and a heat flow.
Observe that in general the state for each reservoir at times $t>0$
will not be the invariant state at a given temperature which is
stationary for the isolated reservoir. It is this fact which makes the
problem of general Hamiltonian reservoirs much more difficult that
than of stochastic reservoirs. It is only in very special cases
(essentially no interaction inside the reservoir) such as that
discussed in subsection \ref{HamSto} where this can be dealt with.

For thermostated systems the temperature of the reservoirs is already
given by the equations of motion so we expect again to have a unique
invariant distribution. Moreover these systems have the property that
the phase space volume is not conserved by the dynamics so that, in
general, no invariant measure will be absolutely continuous with
respect to Lebesgue measure. In this case we need a criterion to chose
the ``physical'' invariant distribution. A natural choice are the so
called SRB states.  These can be characterized by assuming that the
system was in equilibrium in the very distant past and that at some
point a forcing was switched on and this drove the system to a steady
state distribution. More mathematically this means that we consider
the weak limit of a probability distribution absolutely continuous
with respect to Lebesgue, \eg the canonical distribution that
characterize the system when all thermostats have the same
temperature, under the time evolution, \ie if $\Phi^t$ is the flow
describing the evolution of the thermostated system then a state $\mu$
is called SRB if it has the property that
\begin{equation}\label{adj}
\mu_{SRB}(dX)=\lim_{t\to\infty}\bar\Phi^t\lambda(dX)\,,\qquad X=(P,Q)
\end{equation}
where $\bar\Phi^t$ indicates the adjoint and $\lambda(dX)$ is the
given initial distribution. We observe that we often cannot choose
directly the Lebesgue distribution because the phase space of our
system in not compact like for the Nos\'e-Hoover thermostat
Eq.(\ref{nh2}).

Although the definition Eq.(\ref{adj}) is interesting for its
similarity to the ones used in the previous system+reservoirs models
another characterization of these measure is obtained by saying that
they represent the statistics of the motion. More precisely, given any
observable, the average of this observable with respect to the SRB
distribution is equal to its time average along a trajectory starting
from almost every point (with respect to Lebesgue measure).  In
formulae we can say that if $\mu$ is the SRB distribution then
\begin{equation}\label{SRB}
\lim_{t\rightarrow \infty} \frac{1}{t} \int_0^t dt  \delta_{\Phi^t(X)}\,=\,
\mu_{SRB}(dX)
\end{equation}
for a set of $X$ of full (or at least positive) Lebesgue measure.  In
Eq.(\ref{SRB}) the limit is to be understood as a weak
limit\footnote{The only thermostated ``physical'' model for which a
SNS corresponding to an SRB measure with the desired properties has
been proven is the Moran-Hoover model of a single particle moving
among fixed periodic scatterers (see Section \ref{sec:intro}) to which is
added an external electric field and a Gaussian thermostat\cite{CELS}.}.  

\subsection{Heat Flow and Entropy Production in Reservoirs.}\label{flow}

Since our interest here is specifically in Fourier's law our next
question about the stationary state of the system coupled to two
reservoir at different temperatures is the existence and nature of the
heat flux across the system. It is clear that even existence is not
automatic: most trivially just imagine that our system is composed of
two noninteracting parts each coupled to a single reservoir.

To study the heat flux through the system we first define a local
energy density. For a fluid let $\bq\in\Lambda\subset\RRR^d$ then
\begin{equation}
h(\bq;P,Q)=\sum_{i=1}^{N}\delta(\bq-\bq_i)\left[{{\bp_i}^2 \over 2m} +
\sum_{j\ne i} \phi(\bq_j-\bq_i) + u(\bq_i)\right]\label{localen_f}
\end{equation}
where the square bracket is identical to that in Eq.(\ref{loceq}).
For a crystal with nearest neighbor interactions we define the local
energy density at site $\bi$ as
\begin{equation}
h(\bi;P,Q)=\frac{\bp_\bi^2}{2m}+U(\bq_\bi)+\frac{1}{2d}\sum_k
\left(V(\bq_\bi-\bq_{\bi-1_k})+
V(\bq_{\bi+1_k}-\bq_\bi)\right)\label{localen}
\end{equation}
where $1_k$ is the $d$-dimensional vector with all components 0
except the $k$-th equal to 1.  

Given the local energy density we can define a local microscopic heat
flow $\Psi$ through the continuity equation. To avoid repetition we
shall do so only for the crystal. Writing
\begin{equation}
\frac{d h(\bi;P,Q)}{dt}=\tilde\nabla\Psi(\bi)\label{continuity}
\end{equation}
where $\tilde\nabla\Psi(\bi)=\sum_k\tilde\partial_{i_k}\Psi_k(\bi)$
with $\tilde\partial_{i_k}\Psi_j(\bi)=(\Psi_j(\bi+1_k)-
\Psi_j(\bi))/2$.  It easy to verify that
\begin{equation}\label{flux}
\Psi_k(\bi)={\bf f}(\bq_\bi-\bq_{\bi+1_k})\frac{\bp_\bi+\bp_{\bi+1_k}}{2}
\end{equation}
where ${\bf f}(\bq)=- \nabla V(\bq)$.
We will usually be interested in the heat flow $\Phi(j)$ through the
plane $\{\bi \in \Lambda\,;\, i_1=j\}$. It is clear that we can
integrate eq.(\ref{continuity}) and obtain
\begin{equation}
\Phi(j)\,=\, \sum_{ \bi=(j,i_2,\dots,i_d), \atop 1 \le i_l \le N_l }
\Psi_1(\bi).
\end{equation}

Of course the heat current inside the system in the steady state is
just the energy flux from one reservoir to the other, presumably from
the one with the higher temperature to the one with the lower one.
Let $S^t$ be the time evolution for observables (averaged over the
realizations if the evolution is stochastic). Since the equations of
motion are Hamiltonian except at the boundary, one finds that the time
derivative of the energy is given by
\begin{equation}
\frac{d}{dt} S^t H(P,Q)\,=\, -S^t\left(\Phi_L + \Phi_R \right)
\end{equation}
where $\Phi_L$ ($\Phi_R$) depends only on the variables of the left
(right) boundary of the system. It is natural to interpret $\Phi_L$
as the flow of energy from the system to the left heat reservoir and
similarly for $\Phi_R$. We suppose that  a stationary state $\mu$ exists, 
and, for any observable $f$, we set $\mu(f) = \int f d\mu$. One obtains
\begin{equation}
- \mu \left(\Phi_L + \Phi_R \right) \,=\, \mu \left(\frac{d}{dt} S^t
H(p,q) \right) \,=\, 0\,,
\end{equation}
and therefore 
\begin{equation}
\mu (\Phi_L)\,=\, - \mu (\Phi_R)\,.
\end{equation} 

To check that the heat flux is indeed as expected it is useful to
define the entropy production $\sigma$ of the reservoirs as\cite{BeLe}
\begin{equation}
\sigma\,=\,  \frac{\Phi_L}{T_L} + \frac{\Phi_R}{T_R}\,,
\label{sig}
\end{equation}
i.e., $\sigma$ is the sum of the energy flows into the reservoirs
divided by the temperatures of the reservoirs. This (microscopic)
definition of the reservoirs entropy production is in accordance with
our notion of heat reservoir at specified temperature. It does not
require that the system itself be close to equilibrium. A convenient
way of proving that if $T_L>T_R$ then heat is flowing through the
system from left to right is to show that in the steady state
\begin{equation}\label{positive}
\mu(\sigma)\,=\, \left( \frac{1}{T_R} - \frac{1}{T_L} \right) 
\mu(\Phi_R)\, \ge \, 0 \,, 
\end{equation}
and 
\begin{equation}\label{strictpositive}
\mu(\sigma) \,=\, 0 \quad {\rm ~if~and~only~if} \quad T_L=T_R\,. 
\end{equation}
(One may also consider the heat flow $\Phi(i)$ (or the corresponding
$J(\bq)$ for a fluid) inside the system and formally define a
corresponding entropy production $\sigma_i = (T_R^{-1} -
T_L^{-1})\Phi(i)$. One obviously has that $\mu(\Phi_L) = \mu(\Phi_i)$
in the stationary state but $\sigma_i$ is not the macroscopic entropy
production density inside the system\cite{EvMo,CL1,CL2}.)

If $T_L$ and $T_R$ are close, one expects linear response theory to be
valid.  Setting $T=(T_L+T_R)/2$ and $\delta T= (T_L-T_R)$, formal
perturbation theory gives
\begin{equation}
\mu (\sigma) \,=\, \int_0^\infty dt \,\mu_0 (\sigma S_0^t \sigma)  +
{\rm~lower~orders~in~} \delta T\,, 
\label{musigma}
\end{equation}
where $\mu_0$ is the equilibrium Gibbs state at temperature $T$ and
$S_0^t$ the time evolution for observables with $T_L=T_R=T$.  For the
heat flux $\Phi=\Phi(j)$ one obtains
\begin{equation}\label{grku}
\mu (\Phi) \,=\, \frac{T_L-T_R}{T^2} \int_0^\infty dt \,\mu_0 (\Phi
S_0^t \Phi) + {\rm~lower~orders~in~} \delta T \,.
\label{muphi}
\end{equation}
It is important to note that in Eqs.(\ref{musigma}) and (\ref{muphi}),
the reservoirs are still present via the time evolution $S_0^t$.
Although very similar this is not the Green-Kubo formula which will be
discussed below.

\subsection{\em Fourier's Law.}  

Assuming that 1. and 2. have been proved, we can then define the heat
conductivity $\kappa_{\cal L}$ as in Section \ref{sec:intro}, where ${\cal
L}$ the length of the system (fluid or crystal) in microscopic units
and $A$ the area of its crossection. Since $\delta T/{\cal L}$ 
is the average temperature gradient the heat
conductivity at temperature $T$ should then be given by
\begin{equation}\label{refkappa}
\kappa\,=\, \lim_{\LL\rightarrow \infty} \LL  \lim_{\delta T \rightarrow 0}
\frac{1}{\delta T} \left(\mu(\Phi)/A\right)\,,
\end{equation}
i.e., $\kappa=\kappa(T)$ is the heat flux per unit area divided by the
temperature gradient. One might also have taken the limit $A\rightarrow
\infty$ in Eq.(\ref{refkappa}). As might be expected it is the limit
$\LL\rightarrow \infty$ which is the crux of the matter.

\section{Summary of Exact Results}\label{exa} 

We now summarize briefly the limited number of results relating to
points 1-3 of the last section.

\subsection{Fluid systems.}  

Consider a system of $N$ particle in $\Lambda\subset\RRR^d$,
$d\geq 2$, with Hamiltonian given in Eq.(\ref{ham_f}), such that
$u(\bq_i)=0$ and the pair potential $\phi(|\bq|)$ is positive, with
$\phi(0)=C_1$, and $-C_2<\phi'(|\bq|)<0$, $0<C_1,C_2<\infty$, for
$|q|>0$. Then it was shown in \citelow{GoLePr1,GoLePr2,GoKiIa} that,
using Maxwell boundary conditions Eq.(\ref{max}) with temperature
$T(\br)>0$, $\br\in\partial\Lambda$ ($\Lambda$ a regular domain) there
exists a unique stationary $\mu$. Furthermore this $\mu$ is absolutely
continuous with respect to Lebesgue measure on $\Omega=\Lambda^N\times
\RRR^{dN}$ and is approached, as $t\to\infty$, from almost any initial
$(P,Q)$, \ie the set $\Omega'$ for which the approach may fail has
Lebesgue measure zero.

The argument makes use of the boundedness of the force acting on any
particle in the interior of $\Lambda$. This assures that any particle
with a sufficiently high speed will hit the boundary with only little
deviation from a straight path. This and the fact that the force
$-\phi'(|\bq|)$ is everywhere positive insures effective contact
between the system and the stochastic boundaries which, according to
Eq.(\ref{max}), ``spread'' the velocity of particles which hit
it. This yields something like a ``Harris condition'' which guarantees
existence, uniqueness and approach to the stationary state.

Using general technique developed in \citelow{BeLe}
Eq.(\ref{positive}) for systems in contact with stochastic reservoirs
satisfying detailed balance is immediate. It is probably also possible
to prove inequality (\ref{strictpositive}) for such systems but the
latter has not been done generally as far as for we know; see below.

\subsection{Harmonic Crystal.} 

A system with Hamiltonian given by Eq.(\ref{ham}) in which both $V$
and $U_\bi$ (when it does not vanish) are quadratic functions of
their arguments is an ideal harmonic crystal. When such a system is
placed in contact with stochastic reservoirs of the Langevin type the
resulting process and thus also the stationary measure is Gaussian and
one only needs to compute the covariances. This was done essentially
explicitly for a chain in \citelow{RiLeLi}. The most important
difference with the equilibrium state is that there are now
non-vanishing covariances between position and momentum variables
proportional to $\delta T$. One finds uniqueness and approach to the
stationary measure $\mu$ that satisfies Eqs.(\ref{positive}) and
(\ref{strictpositive})

As already mentioned however the heat flux $\mu(\Phi)$ is essentially
independent of $\LL$ and $\kappa_\LL$ defined in Eq.(\ref{refkappa})
grows as $\LL$. (For the case of ``random masses'' $\kappa_\LL$ grows
as $\sqrt{\LL}$\cite{CaLe,OcLe,RiVi}). The solution of
\citelow{RiLeLi} was extended to $d>1$ in \citelow{Na} where there
were also considered various possibilities for $U_\bi$, \eg pinned
down everywhere, only at the boundary or nowhere. Looking at the
invariant measure in the limit $\LL\to\infty$ one finds that the decay
of the position-momentum covariance is rapid when $U_\bi\not=0$ (at
least on the boundary) but does not decay at all when $U_\bi=0$.

The case of an infinite harmonic chain with left and right portions
acting as Hamiltonian reservoirs was investigated in
\citelow{SpLe}. The results are qualitatively the same as for the
stochastic reservoirs: the heat current remain proportional to the
initial temperature difference as $t\to\infty$ and the system
approaches its stationary state which is again a Gaussian
measure\footnote{For an infinite quantum harmonic chain with a
special particle subject to a sufficiently small non harmonic
potential, the existence of stationary states and their mixing
property has been established, both for KMS states~\cite{FiLi,MaGuBo}
and for SNS~\cite{FiLi}.}.

\subsection{Anharmonic Crystals.} 

The anharmonic crystal coupled to Hamiltonian/Stochastic reservoirs
described by Eq.(\ref{epreq}) has been investigated in
\citelow{EPR1,EPR2,EH}. Technical conditions on the growth at infinity
of the potential are needed: either~\cite{EPR1} that ${\cal V}$ is
quadratic at infinity or~\cite{EH} more general polynomial growth. (In
the latter case, the one-body potential $U$ grows more slowly at
infinity then the two-body potential $V$).  One assumes also that the
two-body potential is strictly convex, and this condition alone
implies~\cite{EPR2} that the stationary state is unique.  Under these
conditions the following results hold:
\begin{enumerate}
\item Existence and uniqueness of the stationary state $\mu$.
The stationary state is mixing, \ie any initial distribution will
converges to the stationary state as $t \rightarrow \infty$. The
stationary state has a ${\cal C}^\infty$ density which decays at
infinity at least as fast as a Gibbs state with temperature equal to
the maximum of the temperature of the reservoirs.
\item The stationary state is conducting: One has $\mu(\Phi_R)=0$ if
and only if $T_L=T_R$ and $\mu(\Phi_R)>0$ if $T_L > T_R$. Linear
response theory is valid: For a large class of observables $f$, the
expectation value $\mu(f)$ is a real-analytic function of the
temperature difference $\delta T$. In particular, near equilibrium,
one obtains, with $T=(T_L+T_R)/2$
\begin{equation}
\mu(\Phi_R)\,=\, \frac{\delta T}{T^2}D + O(\delta T^2)\,,
\end{equation}
Eq.\ref{muphi} for the coefficient $D$ has not been proved,
but rather the slightly weaker form
\begin{equation}\label{coeff}
D=  \mu_0 ( \Phi_R  (L_0^{-1} \Phi_R))  \,,
\end{equation}  
where $\mu_0$ is the Gibbs state Eq.(\ref{gg}) with temperature $T$ and
$L_0$ is the generator of the Markovian semi-group $S_0^t$ associated
with the stochastic differential equations (\ref{epreq}) with
$T_L=T_R=T$. Notice that, formally, one has 
\begin{equation}\label{res}
L_0^{-1}=\int_0^\infty dt S_0^t
\end{equation}
and, inserting Eq.(\ref{res}) into Eq.(\ref{coeff}) yields
Eqs.(\ref{muphi}).  In order to prove Eq.(\ref{res}), one needs
presumably some information on the decay of correlations. This has not
been obtained so far. Nothing is known about the dependence of $D$ on
$\LL$ and thus ipso facto about the validity of Fourier's law.

\end{enumerate}

\section{The Green-Kubo Formula}\label{GK}

It is clear that, aside from the case of the harmonic crystal, which
does not satisfy Fourier's law, none of the exact results quoted in
the last section says anything about the local structure, \eg about
local equilibrium in the SNS. This means in particular that at this
time we have no rigorous way of relating the local heat flux
$\mu(\Phi)$ to the gradient of the local temperature as defined in
Eq.(\ref{loceq}). Even in the absence of LTE one can define a local
kinetic temperature by means of the average local kinetic energy. Thus
for the crystal at site $\bi$, $T(\bi)=\mu(\bp_\bi^2/m)/d$. For the
harmonic crystal $T(\bi)$ in found to be uniform away from the ends,
\ie there is no temperature gradient.  Even accepting such a
definition of temperature (in numerical simulation, to be discussed
later, $T(\bi)$ is one of the most directly measured quantities) we
are completely lacking at this point any rigorous or even formal
connection between the $\kappa$ defined in Eq.(\ref{refkappa}) and the
usual Green-Kubo formula for the conductivity which is defined in
terms of the time evolution of an isolated system in
equilibrium. Denoting by ${\hat S}_0^t$ the Hamiltonian evolution of
the isolated system the thermal conductivity $\kappa_{GK}$ is given
by~\cite{KTH,McL}
\begin{equation}\label{realgk}
\kappa_{GK}\,=\, \lim_{L\to \infty} \frac{L}{A}\frac{1}{T^2} 
\int_0^\infty dt \, \langle \Phi {\hat S}_0^t \Phi \rangle 
\end{equation}
where $\langle \cdot \rangle$ denotes the microcanonical average and
the energy density is chosen such that it corresponds to the
thermodynamic energy at temperature $T$ (since we are in equilibrium
this is the same as the kinetic temperature). If the total momentum
${\Pi}$ is conserved it has to be set equal to zero. Alternatively one
may use in Eq.(\ref{realgk}) truncated correlation functions $\langle
\Phi {\hat S}_0^t \Phi \rangle^T_{E,\Pi}= \langle \Phi {\hat S}_0^t
\Phi \rangle_{E,\Pi}-\langle \Phi \rangle_{E,\Pi}^2$ and then average
over $E$ and $\Pi$ using the canonical distribution\footnote{If one
does not fix $\Pi$ in an equilibrium ensemble for which the total
momentum is conserved then the integral in Eq.(\ref{realgk}) is
divergent\cite{PrCa} but this does not say anything about
$\kappa_{GK}$.}. One expects that the equivalence of equilibrium
ensembles will extend also to this case.

The Green-Kubo formula, Eq.(\ref{realgk}), also makes sense for a
system with a few degree of freedom where it can be related to the
variance in the fluctuation of the current; see the article of
Bunimovich and Spohn \cite{BS} for a discussion. There is no clear
connection however between the integral in Eq.(\ref{realgk}) for a
small system and the $\kappa$ in Fourier's law. From a mathematical
point of view it is not even clear how to prove equivalence for
macroscopic systems, \ie show that $\kappa = \kappa_{GK}$.  It would
be nice to find even a formal argument establishing the equivalence.

\section{Entropy Production and Large Deviations}\label{sec:ent_pro}

The proper definition of microscopic entropy production has attracted
much attention in recent years. The interest comes from the
observation of an interesting symmetry property in the large deviation
functional associated to the phase space volume contraction rate in
thermostated systems. This property was first observed numerically in
\citelow{ECM} and then proved under strong hyperbolicity condition in
\citelow{GaCo}. For such systems the phase space volume contraction
has strong connection to the entropy production in Section
\ref{rigstate}. Using this connection the fluctuation theorem has been
extended to large deviations of the entropy production of various
stochastic systems\cite{Kur,LeSp99}.  For crystals with stochastic
reservoirs or for the model considered in Section 4.3, only
formal proofs of the Gallavotti-Cohen fluctuation theorem are
available so far.

The Gallavotti-Cohen fluctuation theorem can be formulated as follows:
In both deterministic and stochastic systems one identifies an
observable $\sigma(X)$ as the ``entropy production'' although its
identification with the thermodynamic entropy production is not always
clear. If one now considers the ergodic average,
\begin{equation}
\sigma_t\left(X\right) = \frac{1}{t} \int_0^t ds \sigma\left(X(s)\right) \,
\end{equation} 
where for deterministic dynamics $X(t)$ is trajectory of the system
while for stochastic dynamics $X(t)$ is a particular realization of
the random process. Then by the large deviation principle (assumed to
hold) there exists an $e(p)$ such that for any interval $I$
\begin{equation}
\lim_{t \rightarrow \infty} \log {\rm Prob}( \sigma_t(x) \in I ) =
\inf_{p \in I} e(p) \,.
\end{equation}
The fluctuation theorem asserts that the odd part of $e(p)$ is
linear with slope $-1$, \ie $e(p)-e(-p)=-p$.

In the deterministic case the entropy production variable that one
consider is, as already mentioned, the phase space contraction rate
which, in the case of Gaussian thermostats, coincides with entropy
production as we defined it in Subsection \ref{flow}. This can be
easily seen by computing the divergence of the equations of motion
quoted at the end of Subsection \ref{Ther}. Already for the case of
the Nos\'e-Hoover thermostat the phase space contraction and the
entropy production are not the same quantities although their average
value in the SNS are the same. For anharmonic chains with a two body
potential given $V(x)= x^2 + \beta x^4$ the validity of the
fluctuation theorem has been checked numerically in \citelow{LeLiPo1}
for the entropy production (but not for the phase space contraction).

The connection between the Gallavotti-Cohen fluctuation theorem and
its various generalizations can be related to the following
observation: For a stochastic model described by a Markov process one
can consider the measure $P$ on the path space induced by the
evolution. Let $\phi$ be a path leading from $X$ to $Y$ in phase space
in the time interval $[0,t]$ and consider now the transformation which
maps the path $\phi$ into the time reversed path leading from $IY$ to
$IX$, where $I$ is the involution on phase space which reverses the
velocities of all particles. This transformation maps the measure $P$
into a new measure ${\overline P}$ which is absolutely continuous with
respect to $P$ with Radon-Nykodym derivative given by
\begin{equation}
\frac{dP}{d {\overline P}} (\phi) \,=\, \exp{\left( R(\phi(t)) -
R (\phi(0)) + \int_0^t \sigma (\phi(s)) ds \right)}\,,
\label{rany}
\end{equation}
where $\sigma$ is the entropy production defined in Eq.(\ref{sig}).
Eq. (\ref{rany}) states, roughly speaking, that the probability of a
time reversed path is equal (up to a boundary term) to the probability
of taht path times the exponential of the integrated entropy production
along this path. This property can be assumed as a general definition
of entropy production\cite{Ma} and can be seen as a generalization of
detailed balance. In fact in the equilibrium case ($T_L=T_R=T$), the
right hand side of Eq.(\ref{rany}) depends on $\phi$ only through its
endpoints and is equal to $\exp{\left( T^{-1} (H(\phi(t)) -
H(\phi(0))) \right)} $ and this is precisely detailed balance.

For thermostated systems the above calculation is quite delicate
because, although one can still consider the measure $P$ on the path
space, it will not be absolutely continuous with respect to the time
reversed measure ${\overline P}$. The original proof of the
fluctuation theorem in \citelow{GaCo} can be thought of as based on
the construction of a series of approximants to the measure $P$ which
satisfies a relation formally similar to Eq.(\ref{rany}).

\section{Local equilibrium}\label{localeq}

Since we have nothing to say about LTE in Hamiltonian heat conducting
systems we discuss briefly some models where the Hamiltonian dynamics
in the bulk is modified by the addition of stochastic forces. These
models show that a mechanism sufficiently strong to destroy the
coherence of the phonons does indeed produce both a temperature
gradient and a normal thermal conductivity, even in one dimensional
systems.

The first such model~\cite{BoRiVi,RiVi} is a harmonic chain coupled to
self-consistent reservoirs. Each particle of the chain is coupled to
its own stochastic reservoir at temperature $T_i$, $i=1,\cdots,N$,
modeled by Langevin dynamics. The reservoirs coupled to the particles
in the bulk supposedly simulate the effect of strong anharmonic
interactions. The temperature of the reservoirs coupled to the first
and last particles are at fixed temperature $T_1=T_L$ and
$T_N=T_R$. The temperature of the remaining reservoirs is fixed by a
self-consistency condition: one requires, that in the stationary state
there is no net energy exchange between the reservoirs and the
particles in the bulk. It is argued~\cite{BoRiVi} that in the limit of
large $N$, the system exhibits a temperature gradient and Fourier's
law is indeed satisfied\cite{RiVi}.

Another model~\cite{Da} is a one-dimensional chain of quantum
mechanical atoms (each of which has a finite number of energy
levels). The first and the last atoms of the chain are each coupled to
heat reservoirs at temperatures $T_L$ and $T_R$. The atoms in the
chains are not directly coupled to each other, but each pair of
nearest neighbors in the chain is coupled to a heat reservoir.  Since
each atom has only a finite number of energy levels, it is possible to
choose the coupling to the intermediate reservoirs in such a way that
no energy is exchanged between the particles in the bulk and the
reservoirs. Compared to the previous model, no self-consistency
condition is needed and the transfer of energy between the
intermediate reservoirs and the system always vanishes, not only in
the stationary state. The model is studied in the weak coupling (or
Van Hove) limit, where the coupling between the chains and all
reservoirs goes to zero and the time is suitably rescaled. In this
limit the evolution is described by a quantum semigroup. For a
suitable choice of couplings the model can be solved exactly and it
exhibits a temperature gradient and Fourier's law.

Another model considered~\cite{KiMaPr} consists of a chain of
uncoupled harmonic oscillators ($U_i(q_i) = q_i^2/2$,
$V(q_i-q_j)=0$). The oscillators at the boundary are coupled to
heat reservoirs modeled by Glauber processes which thermalize the
oscillators according to the Gibbs distribution at temperatures $T_L$
and $T_R$. The energy is exchanged between the oscillators in the
chain according to the following (microcanonical procedure): at each
pair of nearest neighbor sites, there is a clock with exponential law,
when it rings the energy of the pair of particle is redistributed in a
uniform way, keeping the total energy of the system constant. This
model is exactly solvable and satisfies Fourier's law.

\vfill\eject

\section{Numerical results}\label{num_res}

The availability of fast computers has permitted the investigation of
heat current carrying SNS via numerical simulations. Many such
simulations have been performed, see
e.g. \citelow{CaFoViVi,PrRo,LeLiPo1,LeLiPo3,HuLiZh,Ha,AoKu,GLPV} for
recent works on various models.  Earlier works were not always
consistent but recently careful simulations of both one and two
dimensional crystals have become available\cite{LiLi}. A coherent
picture now seems to emerge from the numerical results\cite{LiLi}. We
describe this briefly and refer the reader to \citelow{LeLiPo4} for a
more complete overview.

For obvious reasons, it is convenient in numerical works to keep the
number of degrees of freedom as low as possible. On the other hand
stochastic differential equations, like the ones used in the Langevin
stochastic reservoirs require very good random number generators and
present numerical problems connected with the singularity of the
covariance of the white noise. For this reason the simulations done
using the deterministic thermostat are easier and likely more
reliable. In any case there is good agreement (in the data if not
always in the interpretation) between simulations with different
thermostats.

Since it is not easy to look for invariant measures with numerical
simulations one typically computes the time average of a few
interesting observables along a given trajectory. Assuming the
validity of Eq.(\ref{SRB}) this represents averages with respect to
the invariant measure. This permits investigation of questions like the
temperature profile in a chain in the steady state or the value of the
conductivity for which, as we have seen, there is no analytical result
at this time. We will focus mainly on this last question.

\begin{enumerate}

\item
In one dimension, where most of the simulations have been conducted,
the conductivity, when $U_\bi(\bq)=0$, appears to behave, as a
function of the length $\LL$ of the chain, as $\LL^{\alpha}$ with
$\alpha$ a positive exponent, $\alpha=0.4$ for the anharmonic
chain. On the other hand if $U_\bi(\bq)\not=0$ (typically one consider
$U_\bi(\bq)=\frac{1}{2}\omega^2\bq^2$) one finds a finite conductivity
if some nonlinearity is present in the system (for the linear case
$\alpha=1$\cite{RiLeLi}). In this situation the exact form of the non
linearity seems irrelevant, analogous result are found adding a 4-th
order term to $U(\bq)$ as well as to $V(\bq)$. Similar results are
obtained if one compactifies the configuration space of $Q$, \eg by
considering each $\bq$ as a point on a torus, obtaining what can be
called a chain of rotators.  In all cases the system has a well
defined, approximately linear, temperature profile, although there can
be a finite jump between the temperature of the first and last
oscillator and the temperature of the respective thermostats. Moreover
for one dimensional systems simulations using different kinds of
reservoirs appear to yield similar values of the exponent $\alpha$.
 
\item
Recent simulations seem to show that the conductivity in two
dimensions is logarithmically divergent if $U_\bi(\bq)=0$. Although it is
not easy to see a logarithm is such a situation the simulation in
\citelow{LiLi} strongly suggests this conclusion. Moreover in this
case one can try to compute the thermal conductivity $\kappa_{GK}$ as
given in Eq.(\ref{realgk})\footnote{In this case the equilibrium
ensemble used is $\langle\cdot\rangle_{E,\pi}$ described in the
comments after Eq.(\ref{realgk})}\ and compare it with the $\kappa$
obtained from Eq.(\ref{refkappa}) (without the limit
$\LL\to\infty$). The agreement obtained from the numerical data
support the validity of the equality $\kappa=\kappa_{GK}$. Although we
do not know any results on this direction we believe that adding a
confining $U(\bq)$ to this system will make the conductivity finite.

\item
We further expect that the conductivity in three dimensions will be
finite, with or without any on site potential.
%(arguments to the effect
%that the Green-Kubo formula diverges whenever there is momentum
%conservation, sometimes stated for one-dimension but formally
%applicable in any dimension, do not, in our opinion, use the correct
%ensemble as discussed in Section \ref{GK}, and so prove nothing).

\end{enumerate}

The above picture can be interpreted in term of the Peierls theory
which, as mentioned in Section \ref{micro} relies on umklap processes
to produce a finite conductivity. We also note that in one and two
dimension, there is no stable crystal without an on site potential,
but there is localization in $d\geq 3$. The variance in the deviation
of an atom near the center from its equilibrium position, when the
boundary atoms are tied down, grows like $\LL$ in $d=1$, like $\log
\LL$ in $d=2$ and is finite in $d=3$\cite{BrLiLe}.

\section*{Acknowledgments}
L. R.-B. is supported in part by the Swiss National Science
foundation.  F. B. and J.L were supported by NSF grant DMR-9813268 and
AFOSR grant F49620-98-1-0207. They acknowledge the hospitality of the
IHES where this work was completed.


\begin{thebibliography}{99}

\bibitem{AoKu} K. Aoki and D. Kusnezov, Preprint 1999 {\hfill\break\sf
http://xxx.lanl.gov/ps/chao-dyn/9910015}.

\bibitem{AsMe} N.W. Ashcroft and N.D. Mermin, {\em Solid State
Physics}, Saunders College 1988.

\bibitem{BaFrSi1} V. Bach, J. Froehlich, and I.M. Sigal, 
\Journal{\em Adv. Math.}{137}{205}{1998}.

\bibitem{BaFrSi2} V. Bach, J. Froehlich, and I.M. Sigal, \Journal{\em
Adv. Math.}{137}{299}{1998}.

\bibitem{BaFrSi3} V. Bach, J. Froehlich, and I.M. Sigal, \Journal{
\CMP}{207}{249}{1999}.

%\bibitem{BaFrSiSo} V. Bach, J. Froehlich, I.M. Sigal, and A. Soffer,
%Preprint. (1998). 

\bibitem{BeLe} P.G. Bergman and J.L. Lebowitz, \Journal{\em
Phys. Rev.}{99}{2}{1955}. 

\bibitem{BoRiVi} M. Bolsterli, M. Rich, and W.M. Visscher,
\Journal{\em Phys. Rev. A}{1}{1086}{1970}.

\bibitem{BrLiLe} H.J. Brascamp, E.H. Lieb and J.L. Lebowitz,
\Journal{\em Bulletin of the International Statistic
Institute}{4}{393}{1975}.

\bibitem{Brush} S. Brusch, {\em The kind of motion we call heat},
North Holland, 1976.

\bibitem{BuSi} L. Bunimovich and Ya. G. Sina\u\i,
\Journal{\CMP}{78}{479}{1981}.

\bibitem{BS} L. Bunimovich and H. Spohn,
\Journal{\CMP}{176}{661}{1996}.

\bibitem{CaFoViVi} G. Casati, J. Ford, F.Vivaldi, and W.M.~Visscher,
\Journal{\em \PRL}{52}{1861}{1984}.

\bibitem{CaLe} A. Casher and J.L. Lebowitz,
\Journal{\JMP}{12}{1701}{1971}.

\bibitem{CELS} N.I. Chernov, G.L.  Eyink, J.L. Lebowitz, and
Ya. G. Sina\u\i, \Journal{\CMP}{154}{569}{1993}.

\bibitem{CL1} N.I. Chernov, and J.L. Lebowitz,
\Journal{\PRL}{71}{2831}{1995} 

\bibitem{CL2} N.I. Chernov, and J.L. Lebowitz,
\Journal{\JSP}{86}{953}{1997}. 

\bibitem{Da} E.B. Davies, \Journal{\em \JSP}{18}{161}{1978}.

\bibitem{De} P. Debye, {\em Vortr\"age \"uber die Kinetische Theorie
der W\"arme}, Teubner, 1914.

\bibitem{DeJa} J. Derezinski and V. Jaksic, Preprint 1998. 

%\bibitem{DhDh} A. Dhar and D. Dhar, Preprint 1999. 

\bibitem{EH} J.-P.~Eckmann and M.~Hairer, Preprint 1999 {\hfill\break\sf
http://xxx.lanl.gov/ps/chao-dyn/9909035}.

\bibitem{EPR1} J.-P.~Eckmann, C.-A.~Pillet, and L.~Rey-Bellet,
\Journal{\CMP}{201}{657}{1999}.

\bibitem{EPR2} J.-P.~Eckmann, C.-A.~Pillet, and L.~Rey-Bellet,
\Journal{\JSP}{95}{305}{1999}. 

\bibitem{ELM1} R. Esposito, J.L. Lebowitz and R. Marra, 
\Journal{\CMP}{160}{49}{1994}. 

\bibitem{ELM2} R. Esposito, J.L. Lebowitz and R. Marra, 
\Journal{\JSP}{78}{389}{1995}. 

\bibitem{ELM3} R. Esposito, J.L. Lebowitz and R. Marra, 
\Journal{\JSP}{90}{1129}{1998}. 

\bibitem{ECM} D.~J.~Evans, E.~G.~D.~Cohen, and G.~P.~Morris,
\Journal{\em Phys. Rev. Lett.} {71}{2401}{1993}.

\bibitem{EvMo} D.J.~Evans and G.P.~Morris, {\em Statistical Mechanics
of Nonequilibrium Liquids}, Academic Press, San Diego 1990.

%\bibitem{EyLeSp1} G. Eyink, J.L. Lebowitz and H. Spohn,
%\Journal{\CMP}{132}{253}{1990}.

%\bibitem{EyLeSp2} G. Eyink, J.L. Lebowitz and H. Spohn,
%\Journal{\CMP}{140}{507}{1991}.

%\bibitem{EyLeSp3} G. Eyink, J.L. Lebowitz and H. Spohn,
%\Journal{\JSP}{88}{385}{1996}.

\bibitem{FaGoSp} J. Farmer, S. Goldstein, and E.R. Speer,
\Journal{\JSP}{34}{263}{1984}.

\bibitem{FiLi} F. Fidaleo and C. Liverani \Journal{\JSP}{}{98}{2000}.

%\bibitem{Fo} J. Ford, \Journal{\em Phys. Rep.}{5}{271}{1992}.
 
%\bibitem{FoKaMa} G.W Ford, M. Kac, and P.Mazur 
%\Journal{\em Phys. Rep.}{5}{271}{1992}.

\bibitem{FFL} J. Fritz, T. Fumaki, and J.L. Lebowitz,
\Journal{Probab. Theory Related Fields}{99}{211}{1994}.

\bibitem{GaCo} G.~Gallavotti and E.G.D.~Cohen,
\Journal{\JSP}{80}{931}{1995}.

\bibitem{G1997} G.~Gallavotti, \Journal{\em Physica D}{105}{163}{1997}

\bibitem{G1998} G.~Gallavotti, \Journal{\em Chaos}{8}{384}{1998}

\bibitem{GLPV} C. Giardina, R. Livi, A. Politi, and M. Vassalli, 
Preprint 1999.
  
%\bibitem{Gentile} G. Gentile, \Journal{\em Forum Mathematicum}{10}{89}{1998}.

%\bibitem{GaKiMaPr} A. Galves, C. Kipnis, C. Marchioro, and
%E. Presutti, \Journal{\em \CMP}{81}{127}{1981}.

\bibitem{GoKiIa} S. Goldstein, C. Kipnis and N. Ianiro,
\Journal{\JSP}{41}{915}{1985}.

\bibitem{GoLePr1} S. Goldstein, J.L. Lebowitz and E. Presutti, in {\em
Colloquia Mathematica Societati J\'anos Bolay} 27 (403), 1979.

\bibitem{GoLePr2} S. Goldstein, J.L. Lebowitz and E. Presutti, in {\em
Colloquia Mathematica Societati J\'anos Bolay} 27 (421), 1979.

\bibitem{GoLeRa1} S. Goldstein, J.L. Lebowitz and K. Ravishankar,
\Journal{\CMP}{85}{419}{1982}.

\bibitem{GoLeRa2} S. Goldstein, J.L. Lebowitz and K. Ravishankar,
\Journal{\JSP}{43}{303}{1986}.

\bibitem{Ha} T. Hatano. \Journal{\em Phys. Rev. E}{59}{R1}{1999}.

%\bibitem{Ho1} W.G.~Hoover, \Journal{\em Phys. Rev. A}{31}{1695}{1985}.

\bibitem{Ho} W.G.~Hoover, {\em Computational Statistical Mechanics},
Elsevier 1991

\bibitem{HuLiZh} B. Hu, B. Li and H. Zhao,
\Journal{Phys. Rev. E}{57}{2992}{1998}.

\bibitem{JaPi1} V. Jaksic and C.-A. Pillet,
\Journal{\CMP}{176}{619}{1996}.

\bibitem{JaPi2} V. Jaksic and C.-A. Pillet,
\Journal{\CMP}{178}{627}{1996}.

\bibitem{JaPi3} V. Jaksic and C.-A. Pillet, \Journal{\em Acta
Math.}{181}{245}{1998}.

\bibitem{Ja} E.A. Jackson, \Journal{\em Rocky
Mount. J. Math.}{8}{127}{1978}.

%\bibitem{JaMi} E.A. Jackson and A. D. Mistriotis, \Journal{\em
%J.Phys.: Condens. Matter}{1}{1223}{1989}.

%\bibitem{KaMa} H. Kaburaki and M. Machida, \Journal{\em
%Phys. Lett. A}{181}{85}{1993}.  

\bibitem{KiLa} C. Kipnis and C. Landim, {\em Scaling limits of
interacting particle systems}, Springer, Berlin 1999.

\bibitem{KiMaPr} C. Kipnis, C. Marchioro and
E. Presutti, \Journal{\em \JSP}{27}{65}{1982}.
 
\bibitem{KoSp} A. Komech and H. Spohn, \Journal{\em Nonlinear
Anal.}{33}{13}{1998}.

\bibitem{KoSpKu} A. Komech, H. Spohn and M. Kunze~M \Journal{\em
Comm. Partial Differential Equations}{22}{307}{1997}.

\bibitem{Ku} R. Kubo, \Journal{\em Acta Phys. Austr.,
Suppl.}{10}{301}{1973}.

\bibitem{Kur} J.~Kurchan, \Journal{\em J. Phys. A}{31}{3719}{1998}.

\bibitem{KTH} R. Kubo, M. Toda and N. Hashitsume, {\em Statistical
Physics II}, Springer Series in Solid State Sciences, Vol 31,
Springer, Berlin 1991.

%\bibitem{La} O.E. Lanford III, in {\em Lecture Notes in Physics 20},
%ed. A. Lenard,  
% (Springer, Berlin, 1973).  

%\bibitem{Le1} J.L. Lebowitz, \Journal{\em
%Phys. Rev.}{114}{1192}{1959}. 

\bibitem{Le2} J.L. Lebowitz, 
\Journal{\em Prog. Theor. Phys. Suppl.}{64}{35}{1978}.

\bibitem{LeBe} J.L. Lebowitz and P.G. Bergman, \Journal{\em
Ann. Physics}{1}{1}{1957}.  

%\bibitem{LeOC} J.L. Lebowitz and A.J. O'Conner,
%\Journal{\JMP}{15}{629}{1974}. 
%
\bibitem{LeSp78} J.L. Lebowitz and H. Spohn,
\Journal{\JSP}{19}{633}{1978}.

%\bibitem{LeSp82} J.L. Lebowitz and H. Spohn,
%\Journal{\JSP}{28}{539}{1982}.

%\bibitem{LeSp82-1} J.L. Lebowitz and H. Spohn,
%\Journal{\JSP}{29}{39}{1982}. 

\bibitem{LePrSp} J.L. Lebowitz, E. Presutti and H. Spohn,
\Journal{\JSP}{51}{841}{1988}.

\bibitem{LeSp99} J.L. Lebowitz and H. Spohn,
\Journal{\JSP}{95}{333}{1999}. 

\bibitem{LeLiPo1} S. Lepri, R. Livi, and A. Politi,
\Journal{\PRL}{78}{1896}{1997}. 

%\bibitem{LeLiPo2} S. Lepri, R. Livi, and A. Politi, \Journal{\em 
%Physica D}{119}{140}{1998}.  
 
\bibitem{LeLiPo3} S. Lepri, R. Livi, and A. Politi, \Journal{\em
Europhys. Lett.}{43}{271}{1998}.

\bibitem{LeLiPo4} S. Lepri, R. Livi, and A. Politi,  In preparation. 

\bibitem{LiLi} S. Lippi and R. Livi, Preprint 1999 {\hfill\break\sf
http://xxx.lanl.gov/ps/chao-dyn/9910034}.

\bibitem{LO} C. Liverani and S. Olla, \Journal{\CMP}{189}{481}{1997}

\bibitem{Ma} C.~Maes, \Journal{\em  J. Stat. Phys.}{95}{367}{1999}.

\bibitem{MaGuBo} H. Maassen, M. Guta and D. Botvich, Preprint 1999
{\hfill\break\sf http://www.ma.utexas.edu/mp$\underline{\ }$arch/99/99-220.ps.gz}.

\bibitem{McL} J. A. McLennan, {\em Introduction to nonequilibrium statistical
mechanics}, Prentice Hall, 1989. 

%\bibitem{MoBu} F. Mokross and H. B\"uttner, \Journal{\em J. Phys. C:
%Solid State Phys.}{16}{4539}{1983}. 

\bibitem{Na} H. Nakazawa, \Journal{\em Supplement of the Progress of
Theoretical Physics}{45}{231}{1970}.

\bibitem{No} S.~Nos\'e, \Journal{\em J. Chem. Phys.}{81}{511}{1984}.

\bibitem{OcLe} A.J. O'Connor and  J.L. Lebowitz,
\Journal{\JMP}{15}{692}{1974}.  

\bibitem{OVY} S. Olla, S.R.S. Varahadan, H.-T. Yao,
\Journal{\CMP}{155}{523}{1993}.

\bibitem{Pe1} R.E. Peierls, {\em Quantum Theory of Solids}, University
press 1956. 

\bibitem{Pe2} R.E. Peierls, In {\em Theoretical Physics in the
Twentieth Theory}, Interscience 1960.

\bibitem{PrRo} T. Prosen and M. Robnik, \Journal{\em J. Phys. A:
Math. Gen. }{25}{3449}{1992}. 

\bibitem{PrCa} T. Prosen and D.K. Campbell, Preprint 1999 {\hfill\break\sf
http://xxx.lanl.gov/ps/chao-dyn/9910024}.

\bibitem{RiLeLi} Z. Rieder, J.L. Lebowitz, and E.H. Lieb,
\Journal{\JMP}{8}{1073}{1967}.  

\bibitem{RiVi} M. Rich  and W.M. Visscher,
\Journal{\em Phys. Rev. B}{11}{2164}{1975}.

\bibitem{RuGr} R.J. Rubin and W.L. Greer,  \Journal{\JMP}{12}{1686}{1971}.

%\bibitem{Ru65} D. Ruelle, \Journal{\JMP}{6}{201}{1965}.

%\bibitem{Ru96} D. Ruelle,  \Journal{\JSP}{85}{1}{1996}.

\bibitem{Ru99} D. Ruelle, Preprint 1999 {\hfill\break\sf
http://www.ma.utexas.edu/mp$\underline{\ }$arch/99/99-220.ps.gz}.

\bibitem{Sp} H. Spohn, {\em Large Scale Dynamics of Interacting Particle}, 
Springer Verlag, 1991.

\bibitem{SpLe} H. Spohn and J.L. Lebowitz, \Journal{\CMP}{54}{97}{1977}.

\bibitem{EZ} F. Zhang, D.J. Isbister and D.J. Evans, Preprint 1999,
{\hfill\break\sf http://xxx.lanl.gov/ps/chao-dyn/9910024}.


\end{thebibliography}
\end{document}